\documentclass[useAMS,usenatbib]{mn2e}
\usepackage{tikz}
\usepackage{amssymb}
\usepackage[normalem]{ulem}

\newcommand{\Msolar}{{\rm M_{\odot}}}   
\newcommand{\Qmin}{Q_{\rm min}}  
\newcommand{\betacrit}{\beta_{\rm crit}}  

\title[Resolution effects on fragmentation]{Non-convergence of the critical cooling timescale for fragmentation of self-gravitating discs}
\author[Farzana Meru and Matthew R. Bate]{Farzana Meru$^{1,2}$\thanks{farzana@astro.ex.ac.uk} and Matthew R. Bate$^1$\\
$^1$School of Physics, University of Exeter, Stocker Road, Exeter, EX4 4QL, UK\\
$^2$Institut f\"ur Astronomie und Astrophysik, Universit\"at T\"ubingen, Auf der Morgenstelle 10, 72076 T\"ubingen, Germany}
\begin{document}

\maketitle

\label{firstpage}

\begin{abstract}

We carry out a resolution study on the fragmentation boundary of self-gravitating discs. We perform three-dimensional Smoothed Particle Hydrodynamics simulations of discs to determine whether the critical value of the cooling timescale in units of the orbital timescale, $\beta_{\rm crit}$, converges with increasing resolution.  Using particle numbers ranging from 31,250 to 16 million (the highest resolution simulations to date) we do not find convergence.  Instead, fragmentation occurs for longer cooling
timescales as the resolution is increased.  These results suggest that at the very least, the critical value of the cooling timescale is longer than previously thought.  However, the absence of convergence also raises the question of whether or not a critical value exists.  In light of these results, we caution against using cooling timescale or gravitational stress arguments to deduce whether gravitational instability may or may not have been the formation mechanism for observed planetary systems.
\end{abstract}


\section{Introduction}
\label{sec:intro}

There are two quantities that have historically been used to determine whether a self-gravitating disc is likely to fragment.  The first is the stability parameter \citep{Toomre_stability1964},

\begin{equation}
  \label{eq:Toomre}
  Q=\frac{c_{\rm s}\kappa_{\rm ep}}{\pi\Sigma G},
\end{equation}
where $c_{\rm s}$ is the sound speed in the disc, $\kappa_{\rm ep}$ is the epicyclic frequency, which for Keplerian discs is approximately equal to the angular frequency, $\Omega$, $\Sigma$ is the surface mass density and $G$ is the gravitational constant.  \cite{Toomre_stability1964} showed that for an infinitesimally thin disc to fragment, the stability parameter must be less than a critical value, $Q_{\rm crit} \approx 1$.
 
\cite{Gammie_betacool} showed that in addition to the stability criterion above, the disc must cool at a fast enough rate.  Using shearing sheet simulations, he showed that if the cooling timescale can be parametrized as

\begin{equation}
  \label{eq:beta}
  \beta = t_{\rm cool}\Omega,
\end{equation}
where

\begin{equation}
  \label{eq:tcool}
  t_{\rm cool} = u \Big(\frac{{\rm d}u_{\rm cool}}{{\rm d}t}\Big)^{-1},
\end{equation}
$u$ is the specific internal energy and ${\rm d}u_{\rm cool}/{\rm d}t$ is the total cooling rate, then for fragmentation we require $\beta \lesssim 3$, for a ratio of specific heats $\gamma = 2$ (in two dimensions).  Otherwise, internal heating of the disc due to gravitational instabilities can stablise the disc.  \cite*{Rice_beta_condition} carried out three-dimensional simulations using a Smoothed Particle Hydrodynamics (SPH) code and showed that this cooling parameter is dependent on the equation of state.  They showed that fragmentation can occur if $\beta < \beta_{\rm crit}$ where $\beta_{\rm crit} \approx 6-7$ for discs with $\gamma = 5/3$ and $\beta_{\rm crit} \approx 12-13$ for discs with $\gamma = 7/5$.  \cite{Gammie_betacool} and \cite{Rice_beta_condition} also showed that in a steady state disc where the dominant form of heating is that due to gravitational instabilities, since the gravitational stress in a disc can be linked to the cooling timescale in the disc using

\begin{equation}
  \label{eq:beta_alpha}
  \alpha_{\rm GI} = \frac{4}{9} \frac{1}{\gamma (\gamma - 1)} \frac{1}{\beta},
\end{equation} 
the rapid cooling required for fragmentation can also be interpreted as a maximum gravitational stress that a disc can support without fragmenting, which they showed to be $\alpha_{\rm GI, max} \approx 0.06$.  On the other hand, other authors have carried out simulations that suggest that a single critical value of $\alpha_{\rm GI, max}$ may not necessarily be the case: \cite*{Libby_MSci} showed that the critical value of $\beta$ (below which fragmentation will occur if the stability criterion is met) may depend on the disc's thermal history: if the timescale on which the disc's cooling timescale is decreased is slower than the cooling timescale itself (i.e. a gradual decrease in $\beta$) then the critical value may decrease by up to a factor of 2.  More recently, \cite*{Cossins_opacity_beta} showed that the critical value varies with the temperature dependence of the cooling law.  Given the link between $\beta$ and $\alpha_{\rm GI}$ (equation~\ref{eq:beta_alpha}), this brings $\alpha_{\rm GI, max} \approx 0.06$ into question.

Further inconsistencies also appear in the literature: \cite{Rice_Gammie_confirm} found that for a $0.1 \rm M_{\odot}$ disc with surface mass density profile, $\Sigma \propto R^{-7/4}$, extending to a radius, $R_{\rm out} = 25$~au around a $1\rm M_{\odot}$ star, the disc fragments using $\beta=3$ but not for $\beta=5$, whereas for a disc with mass $M_{\rm disc} = 0.25 \rm M_{\odot}$, the disc fragments for $\beta=5$.  On the other hand, \cite{Rice_beta_condition} suggested that the fragmentation boundary is independent of the disc to star mass ratio.  \cite{Cossins_opacity_beta} carried out a simulation of a self-gravitating disc with surface mass density profile, $\Sigma \propto R^{-3/2}$ (c.f. \cite{Rice_beta_condition} who used $\Sigma \propto R^{-1}$), with ratio of specific heats, $\gamma = 5/3$, and showed that the critical value $\beta_{\rm crit} \approx 4$.  Using equation~\ref{eq:beta_alpha}, this is equivalent to $\alpha_{\rm GI, max} = 0.1$ which also brings the result of  $\alpha_{\rm GI, max} = 0.06$ described above into question.  \cite{Meru_Bate_fragmentation} reconciled these inconsistencies and show that the critical value of the cooling timescale is dependent on $\Sigma R^2/M_{\star}$ i.e. the ratio of the surface mass density at radius, $R$, to the stellar mass, and concluded that the critical value of the cooling time determined by \cite{Rice_beta_condition} is not a general rule.  However, many of their results and the previous results can also potentially be explained if the fragmentation is resolution dependent.

Surprisingly, a proper resolution study that demonstrates convergence of the fragmentation criteria with increasing resolution has not been done.  Most authors who simulate self-gravitating discs consider the resolution criterion set by \cite{Bate_Burkert_resolution} \citep[e.g.][]{Mayer_GI_Sci,Lodato_Rice_original,Stamatellos_no_frag_inside_40AU,Forgan_hybrid}.  Some authors carry out a resolution test (in addition to ensuring the resolution criterion is satisfied) by increasing or decreasing the number of particles by a factor of 2 (e.g. \citealp{2007MNRAS.374..590L,Libby_MSci,Cossins_paper1}).  But large changes in resolution have not been tested.  In an SPH code (used by many of the above mentioned authors), if the number of particles is increased by a factor, $f$, the smoothing length, and hence the resolution, is increased by a factor $f^{\frac{1}{d}}$, where $d$ is the number of dimensions.  \cite{Rice_beta_condition} carried out a resolution test on one of their discs by decreasing the number of particles from 250,000 to 125,000.  They found that the fragmentation results appeared to be unaffected by this.  However, given that the simulations were carried out in three-dimensions, it is important to note that this resolution test was equivalent to decreasing the linear resolution by only $\approx 21\%$, so it is unsurprising that significant differences were not seen.  On the other hand, \cite{Libby_MSci}
carried out a simulation of a fragmenting self-gravitating disc and increased the number of particles from 250,000 to 500,000 and suggested that fragmentation may be affected by resolution.  In addition, the earlier work by \cite{Gammie_betacool} carried out a resolution test for simulations that did not fragment but a resolution test was not carried out on the fragmentation boundary.  Other authors (e.g. \citealp{2007MNRAS.374..590L}; \citealp*{Cossins_paper1}) found that a factor of 2 variation in the particle number did not affect their results.  However, these were also for non-fragmenting discs.

In this letter, we carry out a thorough convergence test of the value of $\beta_{\rm crit}$ required for fragmentation using a disc set up that is typical of the above studies.  Our key result is that we are \emph{unable} to obtain convergence and therefore previous results that make conclusions based on a single critical value of the cooling timescale (and hence a critical value of the gravitational stress) need to be reconsidered.  We describe the numerical method adopted in Section~\ref{sec:numerics}.  We then describe the simulations and the results in Section~\ref{sec:sim_results}.  Finally, we discuss our results and make conclusions in Sections~\ref{sec:disc} and~\ref{sec:conc}, respectively.

\section{Numerical method}
\label{sec:numerics} 

Our simulations are carried out using an SPH code originally developed by \cite{Benz1990} and further developed by \cite*{Bate_Bonnell_Price_sink_ptcls} and \cite{Price_Bate_MHD_h}.  We include the heating effects in the disc due to work done on the gas and artificial viscosity.  The cooling in the disc is taken into account using the cooling parameter, $\beta$ (equation~\ref{eq:beta}), which cools the gas on a timescale given by equation~\ref{eq:tcool}.  To model the shocks in the discs, we use an artificial viscosity \citep{Monaghan_Gingold_art_vis} with SPH~parameters fixed at $(\alpha_{\rm SPH}, \beta_{\rm SPH}) = (0.1, 0.2)$ (as used by \citealp{Rice_beta_condition} and \citealp{Meru_Bate_fragmentation}).  For further details, see \cite{Meru_Bate_fragmentation}.

\section{Simulations \& Results}
\label{sec:sim_results}

\begin{table}
\centering
  {\small
\begin{tabular}{llll}
    \hline
    Simulation name & No of particles & $\beta$ & Fragmented?\\
   \hline
    \hline
    31k-beta2 & 31,250 & 2.0 & Yes\\
    31k-beta2.5 & 31,250 & 2.5 & Yes\\
    31k-beta3 & 31,250 & 3.0 & Yes\\
    31k-beta3.5 & 31,250 & 3.5 & No\\
    31k-beta4 & 31,250 & 4.0 & No\\
    250k-beta5 & 250,000 & 5.0 & Yes\\
    250k-beta5.5 & 250,000 & 5.5 & Yes\\
    250k-beta5.6 & 250,000 & 5.6 & borderline\\
    250k-beta6 & 250,000 & 6.0 & No\\
    250k-beta6.5 & 250,000 & 6.5 & No\\
    250k-beta7 & 250,000 & 7.0 & No\\
    250k-beta7.5 & 250,000 & 7.5 & No\\
    2m-beta5.5 & 2 million & 5.5 & Yes\\
    2m-beta6 & 2 million & 6 & Yes\\
    2m-beta6.5 & 2 million & 6.5 & Yes\\
    2m-beta7 & 2 million & 7 & Yes\\
    2m-beta8 & 2 million & 8 & Yes\\
    2m-beta10 & 2 million & 10 & borderline\\
    2m-beta10.5 & 2 million & 10.5 & borderline\\
    2m-beta11 & 2 million & 11 & No\\
    2m-beta15 & 2 million & 15 & No\\
    16m-beta10 & 16 million & 10 & Yes\\
   16m-beta18 & 16 million & 18 & borderline\\
   \hline
  \end{tabular}
}
  \caption{Table showing the simulations carried out and the key fragmenting results.  The simulations labelled as \emph{borderline} are those that show fragments forming which then shear apart in less than 1 ORP.}
 \label{tab:res_sim}
\end{table}

\begin{figure} 
\centering \includegraphics[width=1.1\columnwidth,angle=-90.0]{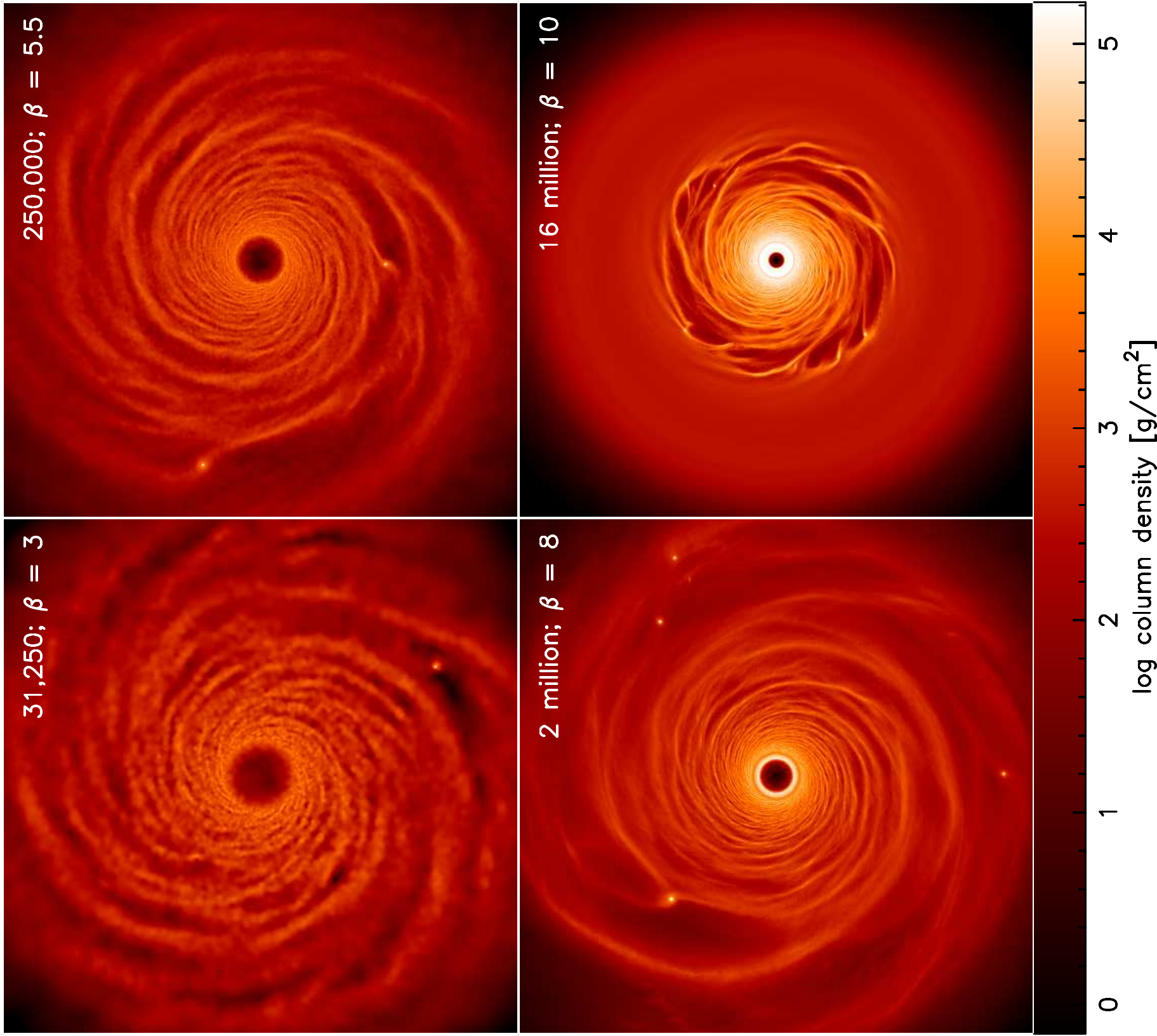}
  \caption{Surface mass density rendered images of the fragmenting discs with 31,250, 250,000, 2 million and 16 million particles (simulations 31k-beta3, 250k-beta5.5, 2m-beta8 and 16m-beta10, respectively).  The images are produced at time $t=5.3$, 6.4, 5.3 and 2.5~ORPs, respectively.  At higher resolution, the disc can fragment for higher values of the cooling timescale.  The axes scale from -25~au to 25~au in both directions.}
 \label{fig:res_frag}
\end{figure}
    
\begin{figure}
\centering \includegraphics[width=1.6\columnwidth,angle=-90.0]{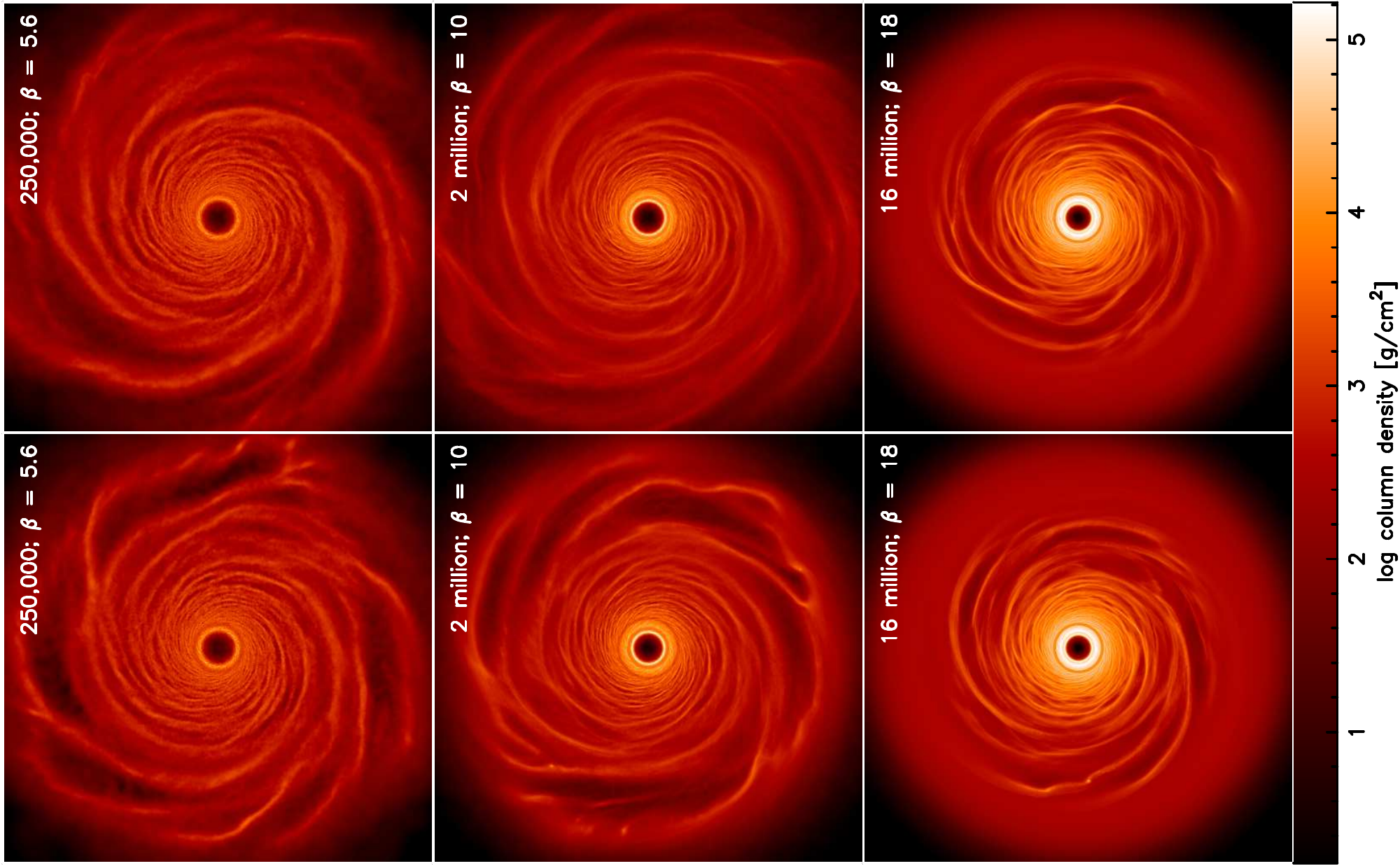}
 \caption{Surface mass density rendered images of the borderline cases with 250,000, 2 million and 16 million particles (simulations 250k-beta5.6, 2m-beta10, 16m-beta18).  The left panels shows a hint of fragmentation at times, $t = 3.8$, 4.8 and 6.0 ORPs (top, middle and bottom panels, respectively).  The right panels shows the equivalent simulations a short time later at times, $t = 4.2$, 5.8 and 6.2 ORPs (top, middle and bottom panels, respectively).  Within 1 ORP, the fragments have been sheared apart, classing these simulations as \emph{borderline}.  The axes scale from -25~au to 25~au in both directions.}
 \label{fig:res_borderline}
\end{figure}

\begin{figure}
\centering
  \includegraphics[width=1.0\columnwidth]{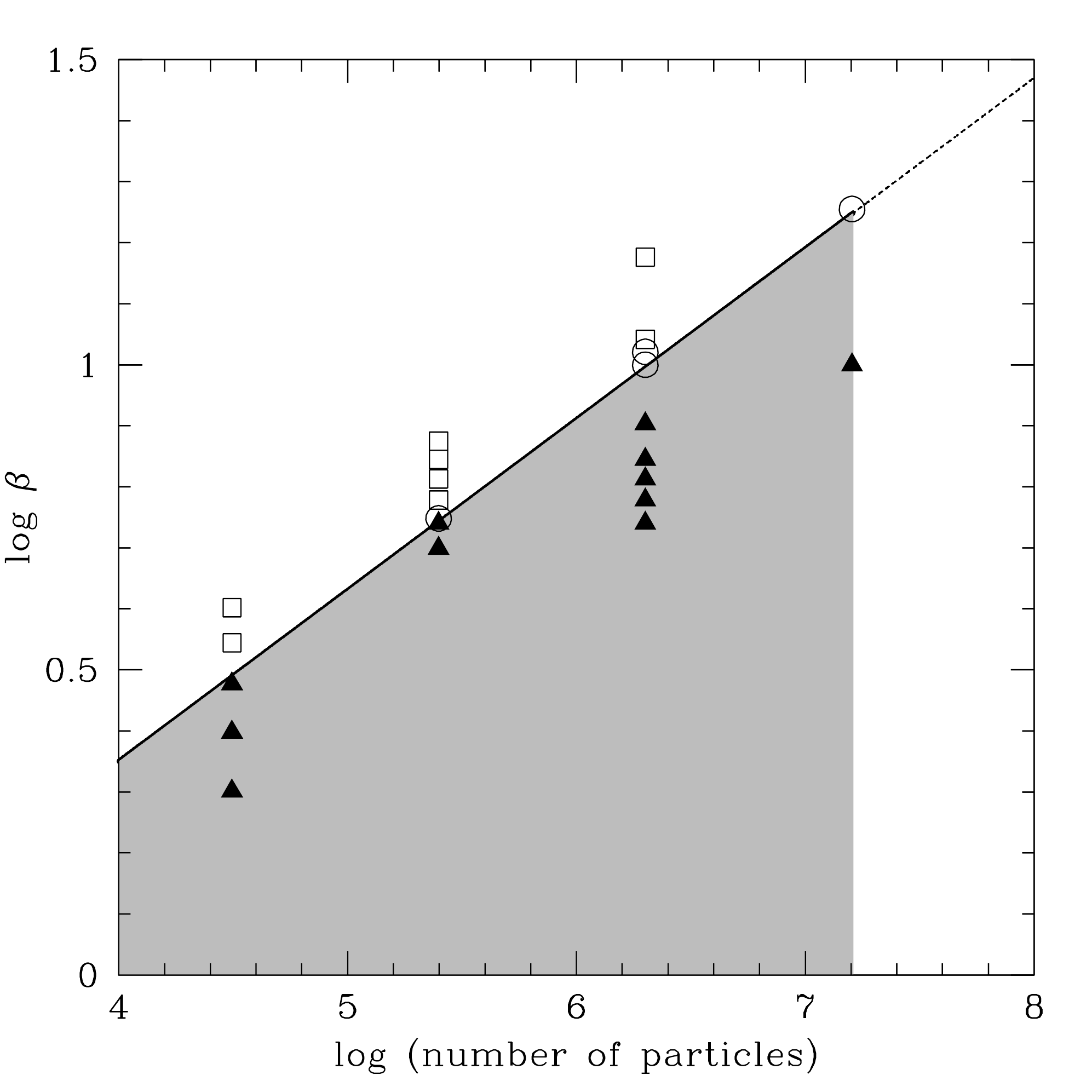}
  \caption{Graph of $\beta$ against resolution of the non-fragmenting (open squares), fragmenting (solid triangles) and borderline (open circles) simulations.  The borderline simulations are those that fragment but whose fragments are sheared apart and no further evidence of fragmentation is seen.  The solid black line shows a dividing line between the fragmenting and non-fragmenting cases and the grey region is where fragmentation can take place.  The graph shows no evidence of convergence of results with increased resolution.  The thin dotted line shows how the trend will continue if convergence is not reached with higher resolution than 16 million particles.  If convergence can be reached, the dotted line is expected to follow a flatter profile.}
 \label{fig:res_beta}
\end{figure}

The disc and star properties used to carry out the simulations in this letter are exactly the same as those used by \cite{Rice_beta_condition}: a $0.1 \Msolar$ disc surrounding a $1 \Msolar$ star, spanning a radial range, $0.25 \le R \le 25$~au.  The initial surface mass density and temperature profiles are $\Sigma \propto R^{-1}$ and $T \propto R^{-1/2}$, respectively, and the temperature is normalised so that the minimum initial Toomre stability value at the outer edge of the disc, $\Qmin=2$.  The discs are modelled with a ratio of specific heats, $\gamma = 5/3$.

\cite{Rice_beta_condition} modelled this disc setup using 250,000 particles.  In Section 4 of \cite{Meru_Bate_fragmentation}, we modelled this same disc setup with $\beta = 5$, 5.5, 5.6 and 6 (Simulations Benchmark1-4 of \citealp{Meru_Bate_fragmentation}) and showed that the critical value of the cooling timescale below which fragmentation occured was $\beta_{\rm crit} \approx 5.6$, a value in reasonable agreement with \cite{Rice_beta_condition}.  We carry out more simulations at this resolution, which we summarise in Table~\ref{tab:res_sim}.  

We also simulate the same disc at higher resolutions by using 2 million and 16 million particles (i.e. the smoothing length is decreased by a factor of 2 and 4, respectively).  In addition, lower resolution testing can also be used to test for convergence.  Therefore, we carry out additional simulations using 31,250 particles (so that the smoothing length is a factor of 2 higher).  We simulate the discs using various values of the cooling timescale in units of the orbital timescale, $\beta$, to determine the critical value, $\beta_{\rm crit}$, at different resolutions.  The simulations were run either for at least 6 outer rotation periods (ORPs) or until the discs fragmented.  Fragments are defined as regions whose surface mass densities are at least two orders of magnitude denser than their surroundings.  Note that there are only two 16 million particle calculations as these are extremely time consuming, each requiring more than 4 months on 64 compute cores of the University of Exeter supercomputer.

Table~\ref{tab:res_sim} summarises the main simulation parameters and the key results in terms of whether the discs fragment or not.  The simulations that show no fragments at all are classed as non-fragmenting discs.  Those discs that fragment and where the fragments remain without being sheared apart are classed as fragmented discs.  There are four simulations that are classed as \emph{borderline} (simulations 250k-beta5.6, 2m-beta10, 2m-beta10.5 and 16m-beta18).  These are discs which show signs of fragmentation but the fragments shear apart rapidly (within 1 ORP) and no further signs of subsequent fragmentation are seen.

Figure~\ref{fig:res_frag} shows images of the fragmenting discs in simulations 31k-beta3, 250k-beta5.5, 2m-beta8 and 16m-beta10 for $\beta$ values of 3, 5.5, 8 and 10 simulated using 31,250, 250,000, 2~million and 16 million particles, respectively.  Figure~\ref{fig:res_borderline} shows images of the borderline discs in simulations 250k-beta5.6, 2m-beta10, 16m-beta18.  The left panels shows the discs appearing to form fragments.  However, within 1~ORP, the fragments shear apart and do not appear to form again.

Figure~\ref{fig:res_beta} plots the results in Table~\ref{tab:res_sim}.  The solid black line divides the fragmenting and non-fragmenting simulations and has been included by eye as a fit to where the boundary lies.  For 250,000, 2 million and 16 million particles, the critical value of the cooling timescale in units of the orbital timescale, $\beta_{\rm crit} \approx 5.6$, 10 and 18, respectively, since these have been identified as \emph{borderline} cases.  For the discs simulated with 31,250 particles, I assume that the critical value is between the fragmenting case of $\beta = 3.0$ and the non-fragmenting case of $\beta = 3.5$ and thus take $\beta_{\rm crit} \approx 3.25$.  It can clearly be seen that with the data that is available, the dividing line between the fragmenting and non fragmenting cases increases linearly with linear resolution and therefore a convergence has not been reached.  Figure~\ref{fig:res_beta} also shows how the trend may continue if convergence does not take place at even higher resolution.

\section{Discussion}
\label{sec:disc}

To date, it has been widely accepted that a single critical value of the gravitational stress determines if a disc will fragment.  However, a thorough convergence test has not been carried out.  The dependence of fragmentation on resolution presented in this letter shows that convergence has clearly not taken place.  This therefore calls into question many of the previous results on the fragmentation boundary \citep[e.g.][]{Gammie_betacool,Rice_Gammie_confirm,Rice_beta_condition,Alexander_ecc_discs,Cossins_opacity_beta,Meru_Bate_fragmentation}, as well as any results that have taken the critical value of the gravitational stress to be a single value of $\alpha_{\rm GI} \approx 0.06$ (or a similarly high value of $\alpha_{\rm GI}$), and based conclusions on these (e.g. \citealp{Clarke2009_analytical,Rafikov_SI,Nero_Bjorkman_GI_analysis,Kratter_runts}).

It is possible that results that are \emph{relative} to each other may still stand.  For example, \cite{Rice_beta_condition} clearly showed that a stiff equation of state requires a faster cooling for fragmentation to occur.  Similarly, \cite{Meru_Bate_fragmentation} presented physical arguments for why the cooling that is required for fragmentation may vary for different surface mass densities and star masses.  However, until convergence is demonstrated, all such results must be treated with caution.

If convergence can be obtained at higher resolution, these results imply that $\betacrit$ can be much higher than previously thought.  In light of the new results presented here, we caution against using cooling timescale or gravitational stress arguments to deduce whether certain planetary systems may or may not have formed by gravitational instability.  \cite{Clarke2009_analytical} produced an analytical model for the structure of a gravitationally unstable disc which is subject to realistic cooling.  She showed that for optically thick discs that are sufficiently low in temperature that they are dominated by ice grains,

\begin{equation}
  \alpha_{\rm GI} = 0.4~\bigg(\frac{R}{100~\rm{au}}\bigg)^{\frac{9}{2}},
\end{equation}
for a disc with interstellar opacities and surrounding a $1 \Msolar$ star, where $R$ is the radius being considered.  This relationship shows that for a maximum value of the gravitational stress, a critical radius, $R_{\rm crit}$, can be found outside of which fragmentation can occur (for a disc with a shallow surface mass density profile).  For example, if $\alpha_{\rm GI,max} \approx 0.07$, then, $R_{\rm crit} \approx 68$~au.  The critical radius also scales as $M_{\star}^{1/3}$ (as described by \citealp{Clarke2009_analytical}) and also depends on the metallicity and grain sizes in the disc (as shown by \citealp{Meru_Bate_opacity}).  One might then use this to compare to observational data and conclude whether a system may or may not have formed via gravitational instability.

Table~\ref{tab:Rcrit_res}, however, shows the equivalent value of $\alpha_{\rm GI,max}$ that is associated with the value of $\beta_{\rm crit}$ identified for each resolution considered in this letter, as well as the critical radius outside of which fragmentation may take place according to \cite{Clarke2009_analytical} for central star masses of $1 \Msolar$ and $1.5 \Msolar$.  It can be seen that the critical radius of fragmentation identified for a disc with 16 million particles is smaller than the value identified for a disc with 250,000 by $> 20 \%$.  This can have profound consequences on the interpretation of observational results.  Table~\ref{tab:Rcrit_res} shows that while the value of $\alpha_{\rm GI,max}$ from simulations using 250,000 particles may cause one to conclude that the outer planet around the $1.5 \Msolar$ A star, HR~8799 \citep{HR8799_metallicity}, may be difficult to form by gravitational instability (since its projected separation according to \citealp{HR8799} is $\approx 68$~au), the results using 16 million particles suggest that this planet could have formed by gravitational instability.

Finally, with the results as they stand, it is possible that convergence will never be obtained regardless of the resolution.  If this is the case, it suggests that the problem may be ill posed.  In other words, it may not be possible for a disc to settle into an equilibrium where there is a balance between heating from gravitational instabilities and a {\it simple} imposed cooling timescale.    We notice that once our simulations have developed significant density structure (but well before fragmentation) the temperature of the gas at a given radius tends to be lower at higher densities.  The cooling rate is an imposed function that only depends on radius from the star and takes no account of the density structure (which is not true of a realistic disc).  Therefore, a possible explanation for the lack of convergence is that while the low-density gas in the disc is being heated by the high-density gas passing through it, the high-density gas has no such internal heating source and continues to cool forming denser structures.  The higher the resolution, the worse such a problem may become.  Clearly more testing needs to be done to definitively determine the reason for the lack of convergence.

\begin{table}
\centering
{\small
\begin{tabular}{lllll}
  \hline
  No of & $\beta_{\rm crit}$ & $\alpha_{\rm GI,max}$ & $R_{\rm crit}$ & $R_{\rm crit}$\\
  particles & & & ($M_{\star} = 1\Msolar$) & ($M_{\star} = 1.5\Msolar$)\\
  \hline
  \hline
  31,250 & 3.25 & 0.12 & 77~au & 88~au\\
  250,000 & 5.6 & 0.07 & 68~au & 78~au\\
  2 million & 10.0 & 0.04 & 60~au & 69~au\\
  16 million & 18.0 & 0.02 & 53~au & 60~au\\
  \hline
\end{tabular}
}
\caption{Table showing how the critical radius of fragmentation according to \citet{Clarke2009_analytical} may be affected for a disc surrounding a 1 and $1.5 \Msolar$ star for the different values of $\beta_{\rm crit}$ identified for discs with 31,250, 250,000, 2 million and 16 million particles.}
\label{tab:Rcrit_res}
\end{table}

\section{Conclusions}
\label{sec:conc}

We present a resolution test of the critical cooling rate required for fragmentation of self-gravitating discs.  We find no evidence of convergence with increasing resolution.  These results certainly show that the critical value of the cooling timescale is longer than previously thought.  However, they also open up the possibility that there may be no value of $\beta$ for which such a disc can avoid fragmentation, given sufficient resolution.  The latter implies that a self-gravitating disc that cools at a rate given by

\begin{equation}
\frac{{\rm d}u}{{\rm d}t}=\frac{u \Omega}{\beta},
\end{equation}
and is only heated by internal dissipation due to gravitational instabilities, may not be able to attain a self-regulated state and will always fragment, regardless of the value of $\beta$.  This re-opens the question of what the criterion for fragmentation of a self-gravitating disc really is and in addition, where in a disc fragmentation can realistically occur.  These results cast some serious doubts on previous conclusions concerning fragmentation of self-gravitating discs and the results presented here need to be considered when making conclusions as to whether observed planetary systems may or may not have formed by gravitational instability.

\section*{Acknowledgments}
We thank Jim Pringle for useful discussions.  We also thank the referee for clarification of the conclusions.  The calculations reported here were performed using the University of Exeter's SGI Altix ICE 8200 supercomputer.  The disc images were produced using SPLASH \citep{SPLASH}.  MRB is grateful for the support of a EURYI Award which also funded FM. This work, conducted as part of the award `The formation of stars and planets: Radiation hydrodynamical and magnetohydrodynamical simulations' made under the European Heads of Research Councils and European Science Foundation EURYI (European Young Investigator) Awards scheme, was supported by funds from the participating organizations of EURYI and the EC Sixth Framework Programme.

\bibliographystyle{mn2e}
\bibliography{allpapers}

\begin{thebibliography}{}

\bibitem[\protect\citeauthoryear{{Alexander}, {Armitage}, {Cuadra} \&
  {Begelman}}{{Alexander} et~al.}{2008}]{Alexander_ecc_discs}
{Alexander} R.~D.,  {Armitage} P.~J.,  {Cuadra} J.,    {Begelman} M.~C.,  2008,
  \apj, 674, 927

\bibitem[\protect\citeauthoryear{{Bate}, {Bonnell} \& {Price}}{{Bate}
  et~al.}{1995}]{Bate_Bonnell_Price_sink_ptcls}
{Bate} M.~R.,  {Bonnell} I.~A.,    {Price} N.~M.,  1995, \mnras, 277, 362

\bibitem[\protect\citeauthoryear{{Bate} \& {Burkert}}{{Bate} \&
  {Burkert}}{1997}]{Bate_Burkert_resolution}
{Bate} M.~R.,  {Burkert} A.,  1997, \mnras, 288, 1060

\bibitem[\protect\citeauthoryear{{Benz}}{{Benz}}{1990}]{Benz1990}
{Benz} W.,  1990, in {J.~R.~Buchler} ed., Numerical Modelling of Nonlinear
  Stellar Pulsations Problems and Prospects, Kluwer, Dordrecht, p.~269

\bibitem[\protect\citeauthoryear{{Clarke}}{{Clarke}}{2009}]{Clarke2009_analyti%
cal}
{Clarke} C.~J.,  2009, \mnras, 396, 1066

\bibitem[\protect\citeauthoryear{{Clarke}, {Harper-Clark} \& {Lodato}}{{Clarke}
  et~al.}{2007}]{Libby_MSci}
{Clarke} C.~J.,  {Harper-Clark} E.,    {Lodato} G.,  2007, \mnras, 381, 1543

\bibitem[\protect\citeauthoryear{{Cossins}, {Lodato} \& {Clarke}}{{Cossins}
  et~al.}{2010}]{Cossins_opacity_beta}
{Cossins} P.,  {Lodato} G.,    {Clarke} C.,  2010, \mnras, 401, 2587

\bibitem[\protect\citeauthoryear{{Cossins}, {Lodato} \& {Clarke}}{{Cossins}
  et~al.}{2009}]{Cossins_paper1}
{Cossins} P.,  {Lodato} G.,    {Clarke} C.~J.,  2009, \mnras, 393, 1157

\bibitem[\protect\citeauthoryear{{Forgan}, {Rice}, {Stamatellos} \&
  {Whitworth}}{{Forgan} et~al.}{2009}]{Forgan_hybrid}
{Forgan} D.,  {Rice} K.,  {Stamatellos} D.,    {Whitworth} A.,  2009, \mnras,
  394, 882

\bibitem[\protect\citeauthoryear{{Gammie}}{{Gammie}}{2001}]{Gammie_betacool}
{Gammie} C.~F.,  2001, \apj, 553, 174

\bibitem[\protect\citeauthoryear{{Gray} \& {Kaye}}{{Gray} \&
  {Kaye}}{1999}]{HR8799_metallicity}
{Gray} R.~O.,  {Kaye} A.~B.,  1999, \aj, 118, 2993

\bibitem[\protect\citeauthoryear{{Kratter}, {Murray-Clay} \&
  {Youdin}}{{Kratter} et~al.}{2010}]{Kratter_runts}
{Kratter} K.~M.,  {Murray-Clay} R.~A.,    {Youdin} A.~N.,  2010, \apj, 710,
  1375

\bibitem[\protect\citeauthoryear{{Lodato}, {Meru}, {Clarke} \& {Rice}}{{Lodato}
  et~al.}{2007}]{2007MNRAS.374..590L}
{Lodato} G.,  {Meru} F.,  {Clarke} C.~J.,    {Rice} W.~K.~M.,  2007, \mnras,
  374, 590

\bibitem[\protect\citeauthoryear{{Lodato} \& {Rice}}{{Lodato} \&
  {Rice}}{2004}]{Lodato_Rice_original}
{Lodato} G.,  {Rice} W.~K.~M.,  2004, \mnras, 351, 630

\bibitem[\protect\citeauthoryear{{Marois}, {Macintosh}, {Barman}, {Zuckerman},
  {Song}, {Patience}, {Lafreni{\`e}re} \& {Doyon}}{{Marois}
  et~al.}{2008}]{HR8799}
{Marois} C.,  {Macintosh} B.,  {Barman} T.,  {Zuckerman} B.,  {Song} I.,
  {Patience} J.,  {Lafreni{\`e}re} D.,    {Doyon} R.,  2008, Science, 322, 1348

\bibitem[\protect\citeauthoryear{{Mayer}, {Quinn}, {Wadsley} \&
  {Stadel}}{{Mayer} et~al.}{2002}]{Mayer_GI_Sci}
{Mayer} L.,  {Quinn} T.,  {Wadsley} J.,    {Stadel} J.,  2002, Science, 298,
  1756

\bibitem[\protect\citeauthoryear{{Meru} \& {Bate}}{{Meru} \&
  {Bate}}{2010a}]{Meru_Bate_opacity}
{Meru} F.,  {Bate} M.~R.,  2010a, \mnras, 406, 2279

\bibitem[\protect\citeauthoryear{{Meru} \& {Bate}}{{Meru} \&
  {Bate}}{2010b}]{Meru_Bate_fragmentation}
{Meru} F.,  {Bate} M.~R.,  2010b, \mnras, pp 1504--+

\bibitem[\protect\citeauthoryear{{Monaghan} \& {Gingold}}{{Monaghan} \&
  {Gingold}}{1983}]{Monaghan_Gingold_art_vis}
{Monaghan} J.~J.,  {Gingold} R.~A.,  1983, Journal of Computational Physics,
  52, 374

\bibitem[\protect\citeauthoryear{{Nero} \& {Bjorkman}}{{Nero} \&
  {Bjorkman}}{2009}]{Nero_Bjorkman_GI_analysis}
{Nero} D.,  {Bjorkman} J.~E.,  2009, \apjl, 702, L163

\bibitem[\protect\citeauthoryear{{Price}}{{Price}}{2007}]{SPLASH}
{Price} D.~J.,  2007, Publications of the Astronomical Society of Australia,
  24, 159

\bibitem[\protect\citeauthoryear{{Price} \& {Bate}}{{Price} \&
  {Bate}}{2007}]{Price_Bate_MHD_h}
{Price} D.~J.,  {Bate} M.~R.,  2007, \mnras, 377, 77

\bibitem[\protect\citeauthoryear{{Rafikov}}{{Rafikov}}{2009}]{Rafikov_SI}
{Rafikov} R.~R.,  2009, \apj, 704, 281

\bibitem[\protect\citeauthoryear{{Rice}, {Armitage}, {Bate} \&
  {Bonnell}}{{Rice} et~al.}{2003}]{Rice_Gammie_confirm}
{Rice} W.~K.~M.,  {Armitage} P.~J.,  {Bate} M.~R.,    {Bonnell} I.~A.,  2003,
  \mnras, 339, 1025

\bibitem[\protect\citeauthoryear{{Rice}, {Lodato} \& {Armitage}}{{Rice}
  et~al.}{2005}]{Rice_beta_condition}
{Rice} W.~K.~M.,  {Lodato} G.,    {Armitage} P.~J.,  2005, \mnras, 364, L56

\bibitem[\protect\citeauthoryear{{Stamatellos} \& {Whitworth}}{{Stamatellos} \&
  {Whitworth}}{2008}]{Stamatellos_no_frag_inside_40AU}
{Stamatellos} D.,  {Whitworth} A.~P.,  2008, \aap, 480, 879

\bibitem[\protect\citeauthoryear{{Toomre}}{{Toomre}}{1964}]{Toomre_stability19%
64}
{Toomre} A.,  1964, \apj, 139, 1217

\end{thebibliography}

\end{document}